\documentstyle[aps,prd,multicol]{revtex}
\tighten

\begin{document}

\title{Spontaneous parity breaking and  fermion masses}

\author
{J. A. Martins Sim\~oes \thanks{simoes@if.ufrj.br}\,, Y. A. Coutinho and C. M. Porto\\ 
Instituto de F\'{\i}sica, Universidade Federal do Rio de Janeiro, RJ, Brazil and \\
Universidade Federal Rural do Rio de Janeiro, RJ, Brazil}

\maketitle
\begin{abstract}
\par
We present a $SU(2)_{L}\otimes SU(2)_{R}\otimes U(1)$ left-right symmetric model for  elementary particles and their connection with the fermion mass spectrum.  New mirror fermions and a minimal set of Higgs particles are proposed. The model can accommodate a consistent pattern for charged and neutral fermion masses as well as neutrino oscillations. The connection between the left and right sectors can be done by the neutral vector gauge boson $Z$ and a new heavy $Z'$.
\par
\end{abstract}
\vskip 1cm
Presented by J.A. Martins Sim\~oes at the International Europhysics Conference on High Energy Physics,\\ July 12-18 2001, Budapest.
\vskip 1cm
\begin{multicols}{2}
 One possible way to understand the origin of the left-right asymmetry in weak interactions is to enlarge the standard model into  a left-right symmetric structure and then, by some spontaneously broken mechanism, to recover  the low energy asymmetric world. 
Left-right models starting from the gauge group \break
$SU(2)_{L}\otimes SU(2)_{R}\otimes U(1)_{B-L}$ were developed by many authors \cite{JCP} and are well known to be consistent with the standard  $SU(2)_{L}\otimes U(1)_Y$. However, for the fermion mass spectrum there is no unique choice of the Higgs sector that can reproduce the observed values for both charged and neutral fermions, neither the fundamental fermionic representation is uniquely defined. 
\par
  As the parity asymmetry is clearly displayed in the electroweak sector of the standard model, we take \cite{COU} as a minimal gauge sector the left-right symmetric group \break
$SU(2)_{L}\otimes SU(2)_{R}\otimes U(1)_{B-L}$, with generators $(T,T^*,Y)$. This group can be considered as a sub-group of many unification groups like the superstring inspired $E_8\otimes E_8'$ or the SUSY - $SO(10)\otimes SO(10)'$. We will assume as a starting point,  new mirror fermions with the following assignment:

\begin{equation}
{\ell_L={\nu \choose e}_L ,\ \nu_R ,\ e_R
 \buildrel {\rm P} \over \longleftrightarrow  L_R={N \choose E}_R,\ N_L,\ E_L}
\end{equation}

\par
The other lepton and quark families follow a similar pattern. This fundamental representation clearly is anomaly free. In this model the parity operation transforms the $SU(2)_{L}\buildrel {\rm P} \over \longleftrightarrow SU(2)_{R}$ sectors, including the vector gauge bosons. This symmetry property can be used to reduce the three group constants $g_L;g_R;g'$ to only two, with $g_L=g_R$. For the other leptonic and quark families a similar structure is proposed. The charge generator is given by $Q = T_3+T_3^*+Y/2$.
\par
In order to break $SU(2)_{L}\otimes SU(2)_{R}\otimes U(1)_Y$ down to $U(1)_{em}$ we introduce two Higgs doublets that under parity are transformed as $\chi_L \leftrightarrow  \chi_R$. Their quantum numbers are (1/2,0,1) and (0,1/2,1) respectively, with the corresponding  vacuum parameters $v_L$ and $v_R$. The model also includes a Higgs field $\Phi$ in the mixed representation (1/2,1/2,0) and new neutral singlets $S$ in  (0,0,0), broken at $s_{GUT}$ or at lower scales. The vacuum parameters are $k$ and $k'$ for the $\Phi$ and $s$ for $S$.
\par 
In the standard model, Higgs doublets are responsible for the gauge boson masses as well as for the fermion masses. For the fermion masses the standard model requires adjusting by hand the Yukawa couplings in order to reproduce the observed mass spectrum. Although this procedure is consistent in the sense that all couplings satisfy $g_i < 1$, there is no experimental confirmation for this hypothesis. One of the main points of our work is to show that Higgs singlets in left-right models can give a more natural charged and neutral fermion mass spectrum. 
\par
The basic physical assumptions of our model are the following. The consistency with the lower energy standard model sector $SU_L(2)\otimes U(1)$ is achieved if we suppose  the symmetry breaking hierarchy $s_{GUT}>>v_R>>v_L>>k,k'$. This was fully developed in ref. [2] and the main result is that we must have the bound  $v_R > 30 \, v_L$. The second assumption is that we must have some see-saw mechanism in the neutrino sector in order to have small neutrino masses. The third point is that the smallness of the electron ( and other charged lepton masses) must also be naturally explained and not adjusted by hand as it is the case for the standard model. As we will shown bellow this can be easily done in models with both left and right handed singlet fermionic  states. This point is the fundamental reason for our choice of the particular fermionic representation given in equation (1). 
\par 

The fermion  mass spectrum depends both on the Higgs choice of the model and on the fundamental fermionic representation. A particular property of the model is the presence of left and right handed singlets in the fundamental representation. This means that we can add to the mass lagrangian new bare terms or new Higgs singlets which have no consequences on the vector gauge boson masses. We consider two new Higgs singlets, one that is coupled to Dirac terms in the mass lagrangian - $S_D$ - and the other that couples to Majorana terms - $S_M$. After spontaneous symmetry breaking they develop vacuum parameters $s_D$ and $s_M$, respectively.
\par
The most general Yukawa lagrangian is given by,
\begin{eqnarray}
{\cal L} & = & f \lbrace \overline{\ell_L} \chi_L \nu_R + \overline{L_R} \chi_R N_L \rbrace \nonumber\\  
& + & f' \lbrace \overline{\ell_L} \chi_L N_L^C + \overline{L_R} \chi_R \nu_R^C \rbrace \nonumber\\ 
& + &
gS_M \lbrace \overline{N_L^C} N_L + \overline{\nu_R^C}\nu_R \rbrace \nonumber \\
& + & g'S_D \lbrace \overline{\nu_R}N_L \rbrace + g" S_D \overline {e_R} E_L\nonumber\\ 
& + & f" \lbrace \overline{\ell_L}\Phi L_R \rbrace  + h.c. 
\end{eqnarray}

For the neutral fermions, the lagrangian after symmetry breaking is 
\vskip 1cm
\begin{eqnarray}
{\cal L} & = & f v_L \overline{\nu_L} \nu_R + f'v_L\overline{\nu_L^c} N_L + fv_R\overline{N_R}N_L \nonumber \\
& + & f'v_R \overline{N_R^C}\nu_R  + gs_M\overline{N_L^C}N_L \nonumber \\
& + & gs_M\overline{\nu_R^C}\nu_R + g' s_D \overline{\nu_R}N_L + f" k \overline{\nu_L}N_R, 
\end{eqnarray}

\par
For the charged fermions we have a similar lagrangian, 
\begin{eqnarray}
{\cal L} & = & f v_L \overline {e_L} e_R +  fv_R \overline {E_R} E_L  \nonumber \\
& + & g" s_D \overline {e_R} E_L +  f" k \overline {e_L} E_R,
\end{eqnarray}

The generalization for the other families is straightforward. The diagonalization 
is most easily done \cite{TPC} by introducing the self conjugated fields ( with $ i,j=\nu,N $)
$\chi_i  = \psi_{iL}+\psi_{iL}^C $ and $\omega_j  = \psi_{jR} +\psi_{jR}^C $. 

\par
We are considering in this paper that the order of magnitude of the fermionic mass spectrum is given by the symmetry breaking scales and their combinations. This is a departure from the standard model procedure of adjusting couplings to masses.  So we are supposing that all couplings are of order one.

In the basis $(\chi_\nu;\omega_N;\chi_N;\omega_\nu)$ the general neutrino mass matrix is:

$$ M_{\nu ,N}=\pmatrix{0&\displaystyle{k/2}&v_L&\displaystyle{v_L
/2}\cr \displaystyle{k/2}&0&\displaystyle{v_R/2}&v_R\cr  v_L&\displaystyle{v_R/2}&s_M&\displaystyle{{s_D}/2}\cr
\displaystyle{v_L/2}&v_R&\displaystyle{{s_D}/2}&s_M\cr}$$

and the charged fermion mass matrix is:
$$ M_{e,E}=\pmatrix{0&k'&0&v_L\cr
                    k'&0&v_R&0\cr
                    0&v_R&0&s_D\cr
                    v_L&0&s_D&0\cr}$$

\par
For this last case, we recover the Dirac formalism by the standard \cite{TPC} $\pi/4$ rotations over the Majorana fields.

\par
The recent SNO \cite{SNO} results increase the experimental evidence for neutrino oscillation and non-zero masses, but still leaving open some theoretical possibilities. In view of the present experimental situation we will not proceed to  fit  all the neutrino masses and mixings but we will look for solutions which could accommodate neutrinos with masses in the $10^{-2}-10^{-3}$ eV range. We present two possible solutions, with $k=k'=0$, which differ in the choice of the symmetry breaking parameters.
\par

$\bullet$ {\bf Model I}

\par
In this model  the Higgs singlets are both broken at $s_{GUT}$. For neutrinos we have the following masses
\begin{eqnarray}
m_{\nu 1}& = & v_L^2/s_{GUT};
m_{\nu 2} = v_R^2/s_{GUT}  \nonumber\\
m_{N1} & \simeq & m_{N2}\simeq s_{GUT}  \nonumber\\
\end{eqnarray}
and for the charged fermions
\begin{eqnarray}
m_e & = & v_L v_R/s_{GUT}  \nonumber\\
m_E & = & s_{GUT}
\end{eqnarray}
 Using  $ v_L=v_{Fermi}$ and $ s_{GUT}=10^{16}$ GeV, we must have 
$v_R=10^{10}$ in order to obtain the correct value for the electron mass. The first generation mass spectrum  is then given by $ m_{\nu1}\simeq10^{-2}$ eV ; $m_{\nu2}\simeq 10$ TeV ;  
$m_{N1}\simeq m_{N2}\simeq 10^{16}$ GeV; $ m_e \simeq 1$ MeV; $ M_E\simeq 10^{16}$ GeV. The smallness of the electron mass is a consequence of a "see-saw" mass relation given by equation 6. This is a departure for the standard model mechanism for fermion masses. 
There is no mixing between $\nu_1$ and $\nu_2$ and we have an example of a sterile neutrino coming from a new exotic sector. The presently observed neutrino mixing must come from a possible generation mixing in the general neutrino mass lagrangian. Standard charged quarks are also found to be in the MeV mass range, with $m_q = v_L v_R/s_{GUT}$. With this high value for $v_R$ we have no experimental accessible accelerator possibilities for new gauge vector bosons. For the Large Hadron Collider (LHC) it has recently been shown \cite{YAC} that heavy neutrino production is limited to masses of a few hundreds of GeV. Heavy neutrinos with masses in the TeV region can be produced in the next generation of colliders.
\par 
The $v_R$ value consistent with the electron mass is of the order of the Peccei-Quinn symmetry breaking scale \cite{JEK} and we can have a possible explanation of the small value for the $\theta$-angle of the strong CP problem, as shown in references \cite{KSB}.
\par
 In order to generate the other families mass spectrum we have the possibility of enlarging the Higgs singlet sector, postulating one new field for each family. In this last case we must have an hierarchy for neutrino masses. 

\par

$\bullet$ {\bf Model II}

\par
In this model lepton number is  spontaneously broken at a scale $s_M \simeq s_{GUT}$ and the Higgs singlet coupled to Dirac mass terms is allowed to be broken at a lower scale. The fermion mass spectrum is given by
\begin{eqnarray}
m_{\nu 1}& = & v_L^2/s_M ;
m_{\nu 2} =  v_R^2/s_M  \nonumber\\
m_{N_1,N_2} & = & s_M\pm s_D/2  \nonumber\\
m_e & = & v_L v_R/s_D  \nonumber\\
m_E & = & s_D
\end{eqnarray}
If we take $v_L=v_{Fermi}$ and $s_M=s_{GUT}$, then the electron and quark masses allow a solution given by $v_R\simeq 10^4$ GeV and $s_D\simeq 10^{10}$ GeV. Here again we have a "see-saw" mechanism for the electron mass. In this model we could have an experimentally accessible new neutral current. One light neutrino has a mass in the $10^{-2}-10^{-3}$ eV range and the other is in the $1 - 0.1$ eV region. These neutrinos are orthogonal and again we have a new sterile neutrino. The Peccei-Quinn symmetry breaking scale reappears in the Higgs singlet sector. Family replication can be recovered with different Dirac singlets and for only one Majorana mass scale we can have a degenerate neutrino mass spectrum.
\par
The generalization for the other families can be easily extended from the above arguments. However, the mixing angle pattern is not so simple. A first approach is to generalize the see-saw mechanism, with mixing angles given by the mass ratios $ \theta_{mix} \simeq m_{\nu}/m_N $, which are very small numbers. There are many models that avoid such a restriction: the introduction of an arbitrary number of right-handed neutrinos \cite{CEC}; some fine-tuning in the neutrino mass matrix \cite{BUC} and any general singular neutrino mass matrix can disconnect mixing parameters from mass ratios. So we will take mixing angles and neutrino masses as independent parameters. 
\par
In Model II we can have the production of new heavy neutral gauge bosons with mass scales accessible at new hadron colliders. Dilepton production in hadron-hadron collisions give a very clear signal for a new Z'. The Large Hadron Collider (LHC) facilities using proton-proton collisions at $\sqrt s= 14$ TeV, will attain higher Z' masses in the 1-4 TeV region.  For an integrated luminosity at LHC of 100 fb$^{-1}$ and cuts of $E_{\ell}> 20$ GeV and ${|\eta|}< 2,5$ we expect 1000 events for $M_{Z'}=1$ TeV and only one event for $M_{Z'}=4$ TeV. 
\par
In conclusion we have shown\cite{COU} that spontaneously broken parity models with a consistent fermion mass spectrum can be built at intermediate scales ranging from $10^3$ to $10^{10}$ GeV. New mirror fermions are present and can be connected to ordinary fermions through neutral currents. We propose a simple  Higgs sector for the model that allows several physically interesting solutions. The charged fermion masses can be generated by a "see-saw" mass relation, analogous to the neutrino sector. Various scenarios for neutrino masses are possible and a more clear experimental definition on neutrino masses will allow to test their reality. Neutral heavy gauge bosons Z and Z', coupled to ordinary and new mirror fermions, are expected to play a fundamental role in the understanding of the left-right symmetry. This is a departure from other models \cite{KOB} that propose gravitation as the connection between the left and right sectors. Experimental consequences of the model could be found at the next generation 
of colliders.
\par
{\it Acknowledgments:} This work was supported by the
 Brazilian agencies: CNPq, FUJB, FAPERJ and FINEP.

\end{multicols}
\end{document}